\documentclass[fleqn,twoside]{article}
\usepackage{espcrc2}
\usepackage{graphicx}
\usepackage{amsmath,amscd,times}

\newcommand{\Pol}{\mathcal{P}}

\title{Tau lepton polarization in quasielastic neutrino-nucleon scattering}
\author{Konstantin S. Kuzmin%
\address[JINR]{Joint Institute for Nuclear Research,
               RU-141980, Dubna, Moscow region, Russia.}%
\address[ITEP]{Institute for Theoretical and Experimental Physics,
               RU-117218, Moscow, Russia.},
        Vladimir V. Lyubushkin%
\addressmark[JINR]%
\address[IrSU]{Physics Department, Irkutsk State University,
               RU-664003, Irkutsk, Russia.}
        and
        Vadim A. Naumov%
\addressmark[JINR]%
\address[INFN]{Dipartimento di Fisica, Universit\`a degli Studi di Firenze
               and INFN Sezione di Firenze, \\
               I-50019, Sesto Fiorentino (FI), Italy.}}
      
\begin{document}

\begin{abstract}
We derive structure functions for the quasielastic production
of octet baryons in $\nu_{\tau}n$ and $\overline{\nu}_{\tau}p$
interactions and study the polarization of $\tau$ leptons
produced in the $\Delta{Y}=0$ reactions. 
Possible impact of the charged second-class currents is investigated
by adopting a simple phenomenological parametrization for the
nonstandard scalar and tensor nucleon form factors. Our choice of the
unknown parameters is made to satisfy the limits obtained in the
(anti)neutrino scattering experiments and rigid restrictions derived
from the nuclear structure studies.
\vspace{1pc}
\end{abstract}

\maketitle

\section{INTRODUCTION}
\label{sec:Introduction}

The forthcoming success of many experimental projects for exploring
neutrino oscillations, nonstandard neutrino interactions, proton
decay and related phenomena hinges on unambiguous reconstruction of
$\tau$ lepton events generated in neutrino-matter interactions and
detected through secondary particles produced in the $\tau$ decays.
Both the energy and angular distributions of these secondaries
are functionals of the $\tau$ lepton polarization; the latter will
be therefore a substantial input parameter for the data processing
in the future experiments.
Some aspects of the subject have been extensively studied in
several recent papers~\cite{Hagiwara:03,Kuzmin:03,Kuzmin:04,Recent}.

In present work, we study the $\tau$ lepton polarization 
in quasielastic (QE) ${\nu}n$ and $\overline{\nu}p$ collisions
with production of an unpolarized $SU(3)$ octet baryon, taking
care to include contributions induced by the second-class currents
(SCC)~\cite{Wilkinson,Gardner:01}.
Since the standard model contains the first-class currents (FCC) only,
it seems fair to examine the potentially measurable SCC effects.

As the most important particular case, our consideration of course
includes the $\Delta{Y}=0$ reactions $\nu_{\tau}n\to\tau^-p$ and
$\overline{\nu}_{\tau}p\to\tau^+n$ and the numerical examples
are done just for this case (Sect.~\ref{sec:Results}).

\section{QE STRUCTURE FUNCTIONS}
\label{sec:W}

The most general form of the electroweak transition current is
given by \cite{LlewellynSmith:72}
\begin{equation}\label{Current_QE}
J_{\alpha}=\langle B;p'|\widehat{J}_{\alpha}|N;p\rangle=
\overline{u}_B\left(p'\right)\varGamma_\alpha\,u_N(p),
\end{equation}
with the vertex function
\begin{align}\label{Vertex_QE}
\varGamma_\alpha
= &\    \gamma_{\alpha}F_V
      +i\sigma_{\alpha\beta}\frac{q^\beta}{2M}F_M
      + \frac{q_\alpha}{M}F_S \nonumber \\
  &\  + \left(\gamma_{\alpha}F_A
      + \frac{p_\alpha+p'_\alpha}{M}F_T
      + \frac{q_\alpha}{M}F_P\right)\gamma_5
\end{align}
defined through the six, in general complex, form factors
$F_i\left(q^2\right)$: FCC- ($i=V,M,A,P$) and
SCC-induced ($i=S,T$).
Here $p$ and $p'$ are the 4-momenta of the initial nucleon $N$
(with mass $M_N$) and final baryon $B$ (with mass $M_B$),
respectively, $q=p'-p$ and $M=\left(M_N+M_B\right)/2$.

The hadronic tensor can be written as
\begin{equation}\label{HadronicTensor_QE}
W_{\alpha\beta}=C_B\sum_{\text{spins}}
J_{\alpha}J_{\beta}^*\;\delta\left(W^2-M_B^2\right),
\end{equation}
where $C_B$ is a dimensionless factor defined by the specific
reaction (e.g., for the $\Delta Y=0$ reactions,
$C_B=\cos^2\theta_C/4$, where $\theta_C$ is the Cabibbo mixing
angle \cite{LlewellynSmith:72}),
$W^2=\left(p+q\right)^2$, and the sum is over spins of initial
and final hadrons.

Then by applying Eqs.~\eqref{Current_QE}, \eqref{Vertex_QE}
and \eqref{HadronicTensor_QE}, one can find the structure
functions $W_i=W_i\left(q^2,W\right)$ involved into the
well-accepted presentation of the hadronic
tensor~\cite{LlewellynSmith:72}
\begin{align}\label{HadronicTensor}
W_{\alpha\beta}
= &\ -g_{\alpha\beta}W_1
     +\frac{p_{\alpha}p_{\beta}}{M^2}W_2
     -\frac{i\epsilon_{\alpha\beta\gamma\delta}
      p^{\gamma}q^{\delta}}{2M^2}W_3 \nonumber\\
  &\ +\frac{q_{\alpha}q_{\beta}}{M^2}W_4
     +\frac{p_{\alpha}q_{\beta}
     +q_{\alpha}p_{\beta}}{2M^2}W_5 \nonumber\\
  &\ +\frac{i\left(p_{\alpha}q_{\beta}
     -q_{\alpha}p_{\beta}\right)}{2M^2}W_6.
\end{align}
After standard calculations we arrive at
\begin{equation}\label{W_i}
W_i=
4C_BM_NM_B\,\omega_i\left(q^2\right)\delta\left(W^2-M_B^2\right),
\end{equation}
where the functions $\omega_i$ are given by
\begin{equation}\label{omega_i}
\omega_i\left(q^2\right)=\omega_i^0\left(q^2\right)
                  +r\,\omega_i^1\left(q^2\right)
                  +r^2\omega_i^2\left(q^2\right)
\end{equation}
and the 15 \emph{nonzero} coefficient functions $\omega^k_i$ in
\eqref{omega_i} are the bilinear combinations of the form factors:
\begin{gather*}
\begin{aligned}
\omega_1^0 = &\  \left(1+x'\right)\left|F_A\right|^2
                           +x'\left|F_V+F_M\right|^2,                \\
\omega_1^2 = &\  \left|F_V+F_M\right|^2,                             \\
\omega_2^0 = &\  \left|F_A\right|^2+\left|F_V\right|^2
                + x'\left|F_M\right|^2+4x'\left|F_T\right|^2,        \\
\omega_2^1 = &\  4\,\text{Re}\left(F_A^*F_T\right),                  \\
\omega_2^2 = &\  4\left|F_T\right|^2,                                \\
\omega_3^0 = &\ -2\,\text{Re}\left[F_A^*\left(F_V+F_M\right)\right], \\
\omega_4^0 = &\ \left(1+x'\right)\left|\tfrac{1}{2}F_M-F_S\right|^2            
                + x'\left|F_P+F_T\right|^2                           \\
             &\ -\text{Re}\left[\left(F_V^*+F_M^*\right)
                          \left(\tfrac{1}{2}F_M-F_S\right)\right.    \\
             &\           \left.+F_A^*\left(F_P+F_T\right)\right],   \\
\omega_4^1 = &\  \text{Re}\left[\left(F_V^*+F_M^*\right)
                          \left(\tfrac{1}{2}F_M-F_S\right)\right.    \\
             &\           \left.+F_A^*\left(F_P+F_T\right)\right],   \\
\omega_4^2 = &\  \left|F_P+F_T\right|^2,                             \\
\omega_5^0 = &\  \omega_2^0+2\,\text{Re}
                 \left[F_S^*\left(F_V-x'F_M\right)\right.            \\
             &\  \left.-F_T^*\left(F_A-2x'F_P\right)\right],         \\
\omega_5^1 = &\  \omega_2^1+\text{Re}\left[F_M^*\left(F_V+F_M\right)
                +2F_A^*F_P\right],                                   \\
\omega_5^2 = &\  \omega_2^2+4\,\text{Re}\left(F^*_PF_T\right),       \\
\omega_6^0 = &\  2\,\text{Im}\left[F_S^*\left(F_V-x'F_M\right)\right.\\
             &\  \left.+F_T^*\left(F_A-2x'F_P\right)\right],         \\
\omega_6^1 = &\ -   \text{Im}\left(F_M^*F_V+2F_A^*F_P\right),        \\
\omega_6^2 = &\  4\,\text{Im}\left(F^*_P F_T\right);
\end{aligned}                                                        \\
r=\left(M_B-M_N\right)/(2M),
\quad
x'=-q^2/\left(4M^2\right).
\end{gather*}

With the formulas for $W_i$ at hand, one can find the lepton polarization
density matrix
$\boldsymbol{\rho}=\frac{1}{2}\left(1+\boldsymbol{\sigma\Pol}\right)$
and the polarization vector
$\boldsymbol{\Pol}=\left(\Pol_P,\Pol_T,\Pol_L\right)$
by applying the generic relations given in Ref.~\cite{Kuzmin:03}
(after fixing an unphysical phase and putting $W_6=0$, these relations
reduce to those of Hagiwara et al.~\cite{Hagiwara:03}).

In the $r=0$ limit, our results agree with those of Ref.~\cite{Kuzmin:03}
and the differential cross section which follow from the obtained formulas in
the standard model limit ($F_S=F_T=0$) reduces to the recent result of Strumia
and Vissani \cite{Strumia:03} derived for the inverse $\beta$ decay taking
account the proton-neutron mass difference.

Let us note that the traditional parametrization
\eqref{Vertex_QE} is not symmetric relative to transformation 
$F_M\leftrightarrow\gamma_5F_T$. The more symmetric choice,
$\frac{i}{2}\sigma_{\alpha\beta}q^{\beta}F_T'$ instead of
$\left(p+p'\right)_{\alpha}F_T$, would result in the following
redefinition of the axial-vector and tensor form factors:
$F_A \mapsto F_A+r F_T'$ and $F_T \mapsto -2F_T'$.
Clearly, after such a redefinition, the functions $\omega_i$ remain
quadratic in $r$. 

\section{SCC EFFECTS}
\label{sec:Results}

For the numerical implementation, we apply the extended
Gari--Kr\"uempelmann model for the Sachs form factors of
proton and neutron \cite{Lomon:02}.
Specifically we choose the so-called ``GKex(02S)'' fit advocated
by Lomon, which is very close numerically to the ``BBA-2003''
parametrization by Budd et al.~\cite{Budd:03}.
We use the standard dipole parametrization for the axial form factor
$F_A$ with the axial mass $M_A=1$~GeV$/c^2$ and the PCAC inspired
relation between the pseudoscalar form factor $F_P$ and $F_A$
suggested by Llewellyn Smith~\cite{LlewellynSmith:72}.

To get some feeling for how big the SCC effects could be, let us consider
the following \emph{toy} model of the scalar and tensor form factors:
\begin{align}
\label{F_S}
F_S\left(q^2\right)&=\xi_Se^{i\phi_S}F_V(0)\left(1-\frac{q^2}{M^2_S}\right)^{-2}, \\
\label{F_T}
F_T\left(q^2\right)&=\xi_Te^{i\phi_T}F_A(0)\left(1-\frac{q^2}{M^2_T}\right)^{-2}.
\end{align}
The model includes six free parameters, $\xi_{S,T}\ge0$, $\phi_{S,T}$
and $M_{S,T}$ and is a straightforward generalization of the models
adopted by several experimental groups~\cite{SCCnu,Ahrens:88}
to constrain the SCC couplings from the measurements of
$\nu/\overline{\nu}$ scattering.

The strongest 90\% C.L. upper limit on the axial SCC strength
$\xi_T$ has been obtained at the Brookhaven AGS experiment with a
$\overline{\nu}_{\mu}$ beam~\cite{Ahrens:88}
as a function of the ``tensor mass'' $M_T$, assuming CVC ($\xi_S=0$),
$M_A=1.09$ GeV$/c^2$ and the simple dipole form of the electromagnetic
form factors.
The limit ranges from about 0.78 at $M_T=0.5$ GeV$/c^2$ to about 0.11 at
$M_T=1.5$ GeV$/c^2$.

Considering  that the $F_S$ contribution into the QE
cross section is suppressed by $\left(m_\mu/m_p\right)^2\sim0.01$, the
constraint to the vector SCC (violating the CVC principle)
is not so strong. Assuming $\xi_T=0$ and $M_S=1.0$ GeV$/c^2$ yields
$\xi_S<1.8$ (90\%\,C.L.)~\cite{Ahrens:88}.
Although CVC violation is not only of academic interest
(see, e.g., Ref.~\cite{Gardner:01} and references therein),
in this short paper we will have to avoid its further discussion
and concentrate on the axial SCC effects. A few examples of the
nonzero $\xi_S$ impact can be found in our recent paper \cite{Kuzmin:04}.

The choice of the remaining unknown parameters in our numerical patterns
is made to fulfil the BNL-AGS limits~\cite{Ahrens:88} (very sensitive to $M_T$)
and the more robust restrictions on the axial SCC coupling constant from
studies of $\beta$ decay of complex nuclei~\cite{Wilkinson} (almost insensitive
to $M_T$). As a conservative upper limit, we accept $\xi_T<0.1$ for any $M_T$
varying between 0.5 and 1.5 GeV$/c^2$.
The phase $\phi_T$ is not constrained, neither by nuclear structure data
nor by measurements of the unpolarized (anti)neutrino-nucleon cross
sections. But the lepton polarization vector is, in general, quite
sensitive to $\phi_T$, even though the strength parameter $\xi_T$ is small.
So we vary $\phi_T$ between $0^\circ$ and $180^\circ$.

\begin{figure}[htb]
\vskip  2mm
\includegraphics[width=\linewidth]{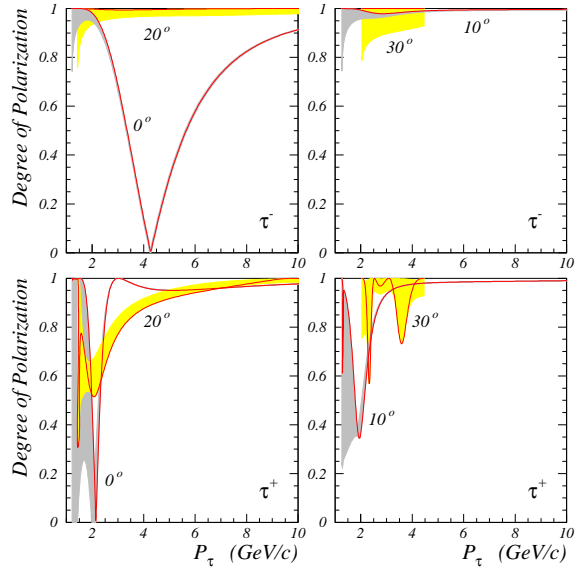}
\vskip -2mm
\caption{\label{fig:P_GK} Degree of polarization,
         $\left|\boldsymbol{\Pol}\right|$, of $\tau$ leptons
         produced in reactions $\nu_{\tau}n\to\tau^-p$ and
         $\overline{\nu}_{\tau}p\to\tau^+n$, as a function of the
         lepton momentum for scattering angles
         $\theta=0^\circ, 10^\circ,20^\circ$ and $30^\circ$.
         The meaning of the curves and filled areas is explained
         in the text.
        }
\end{figure}

\begin{figure}[t]
\vskip  2mm
\includegraphics[width=\linewidth]{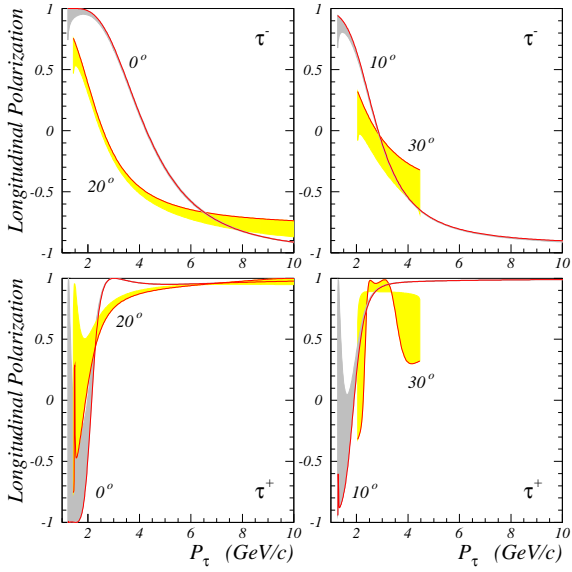}
\vskip -2mm
\caption{\label{fig:P_L_GK} Longitudinal polarization of $\tau^\pm$
         leptons.
         The notation is the same as in Fig.~\ref{fig:P_GK}.
        }
\end{figure}

\begin{figure}[t]
\vskip  2mm
\includegraphics[width=\linewidth]{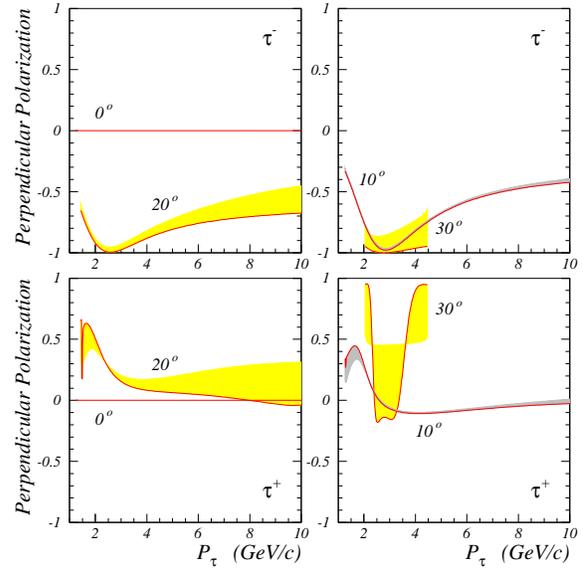}
\vskip -2mm
\caption{\label{fig:P_P_GK} Perpendicular polarization of $\tau^\pm$
         leptons.
         The notation is the same as in Fig.~\ref{fig:P_GK}.
        }
\end{figure}

In figures \ref{fig:P_GK}, \ref{fig:P_L_GK} and \ref{fig:P_P_GK}
we show, respectively, the degree of polarization,
$\left|\boldsymbol{\Pol}\right|=\sqrt{\Pol_P^2+\Pol_T^2+\Pol_L^2}$,
longitudinal polarization, $\Pol_L$, and perpendicular polarization,
$\Pol_P$, of $\tau^\pm$ leptons generated quasielastically in the
$\Delta{Y}=0$ reactions, as functions of the lepton momentum $P_\tau$
(starting from its lowest, kinematically allowed value) for several
scattering angles $\theta$. 
We do not show here the transversal component, $\Pol_T$, which is
nontrivial due to the nonzero phase $\phi_T$ but comparatively small. 
The (anti)neutrino energy is uniquely
defined by the kinematics for each allowed pair $(P_\tau,\theta)$.
The filled areas depict variations of the unknown parameters $\xi_T$,
$M_T$ and $\phi_T$ in Eq.~\eqref{F_T} within the limits
described above and assuming $\xi_S=0$. The curves are for the
standard values of $\left|\boldsymbol{\Pol}\right|$, $\Pol_L$
and $\Pol_P$ calculated with $\xi_T=\xi_S=0$.
 \\
The axial SCC contributions are rather responsive to variation
of each of the parameters $\xi_T$, $M_T$ and $\phi_T$.

As is seen from the figures, the SCC may essentially affect the
polarization vector, particularly at low lepton momenta and large
scattering angles; more sizably in case of $\tau^+$.
However, in the kinematic regions for which the cross section of lepton
production is comparatively large, the SCC effects are not too dramatic
and (especially in case of $\tau^+$) they are more sensitive to small
variations of the \emph{standard} axial and pseudoscalar form factors.
This is a disadvantage for experimental investigation of the SCC
effects but a clear advantage for the future neutrino oscillation
experiments since the relevant uncertainties are not very significant.
Recall that our analysis is only valid within the adopted
\emph{ad hoc} model for the SCC induced tensor form factor,
including somewhat optional range for the tensor mass values.


\section{SUMMARY}
\label{sec:Summary}

We derived the most general formulas for the structure functions
describing the QE production of octet baryons in CC $\nu_{\tau}n$
and $\overline{\nu}_{\tau}p$ interactions; both standard (FCC induced)
and nonstandard (SCC induced) contributions were taken into account.
As an example of application of our result, we studied the axial
SCC effects to the polarization of $\tau$ leptons produced in the
$\Delta{Y}=0$ reactions.

\end{document}